\documentclass[10pt]{iopart}

\usepackage{graphicx}

\begin{document}

\title[Single-Sided Field Free Line Generator Magnet]{Single-Sided Field-Free Line Generator Magnet for Multidimensional Magnetic Particle Imaging}
\author{Alexey Tonyushkin$^{1,2}$}

\address{$^1$Physics Department, University of Massachusetts Boston, Boston, MA 02125 USA}
\address{$^2$ Department of Radiology, Massachusetts General Hospital, Boston, MA 02114 USA}
\ead{alexey.tonyushkin@umb.edu}

\begin{abstract}
Magnetic Particle Imaging (MPI) is an emerging medical imaging modality that is not yet adopted by clinical practice. 
Most of the working MPI prototypes including commercial-grade research MPI scanners utilize cylindrical bores that limit the access to the scanner and the imaging volume. Recently a single-sided or an asymmetric device that is based on a field-free point (FFP) coplanar coil topology has been introduced that shows promise in alleviating access constraint issues. In this paper, we present a simulation study of selection coils for a novel design of a single-sided MPI device that has an advantage of a more sensitive field-free line (FFL) topology.
\end{abstract}

{\it Keywords\/}: magnetic particle imaging, field-free line, single-sided, selection coils
\maketitle
\ioptwocol

\section{Introduction}
Magnetic Particle Imaging (MPI)~\cite{Gleich05,Buzug12} device has yet to be introduced into clinical practice. 
One of the challenges has been the ability upscaling the gradient coils to surround a human body while been able to generate and to drive a sufficiently strong gradient of the magnetic field. These requirements, however, make a demand on a prohibitively high power consumption in the device with the cylindrical geometry.
Therefore alternative topologies such as an open geometry scanner would be highly desirable. One way to make such a practical MPI device is to use a single-sided geometry~\cite{Sattel09}. The single-sided device has all the hardware on one side from the imaging volume and therefore such a device can be used equally on small animals and humans, as well as an MPI spectrometer for medical and material surveys. 
Due to inherent geometrical constrains, {\em i.e.} lack of counteractive coils in the parallel configuration there is a significant challenge in the implementation of such devices. To date, only one type of a single-sided FFP-based device has been demonstrated~\cite{Grafe13}. Here we propose a novel design of Selection-Drive (SD) coils for a single-sided device that is capable of 3-D image encoding. Moreover, different from the recent field-free point (FFP) based developments~\cite{Grafe15,Kaethner15,Grafe16}, in our geometry we utilize a more sensitive field-free line (FFL) configuration~\cite{Weizenecker08,KnoppMedPhys}. The design is capable generating a sufficient magnetic field gradient and drive fields so that the FFL can travel across a volume of a small animal or penetrate deep enough into human organs such as in the vascular or lymphatic systems for sentinel lymph node biopsy~\cite{Grafe12,Shiozawa13}.

\section{Theory}
\subsection{Basic principle of FFL generator}
\label{station}
\begin{figure}[h!]
 \centering
 \includegraphics[width=3.4 in]{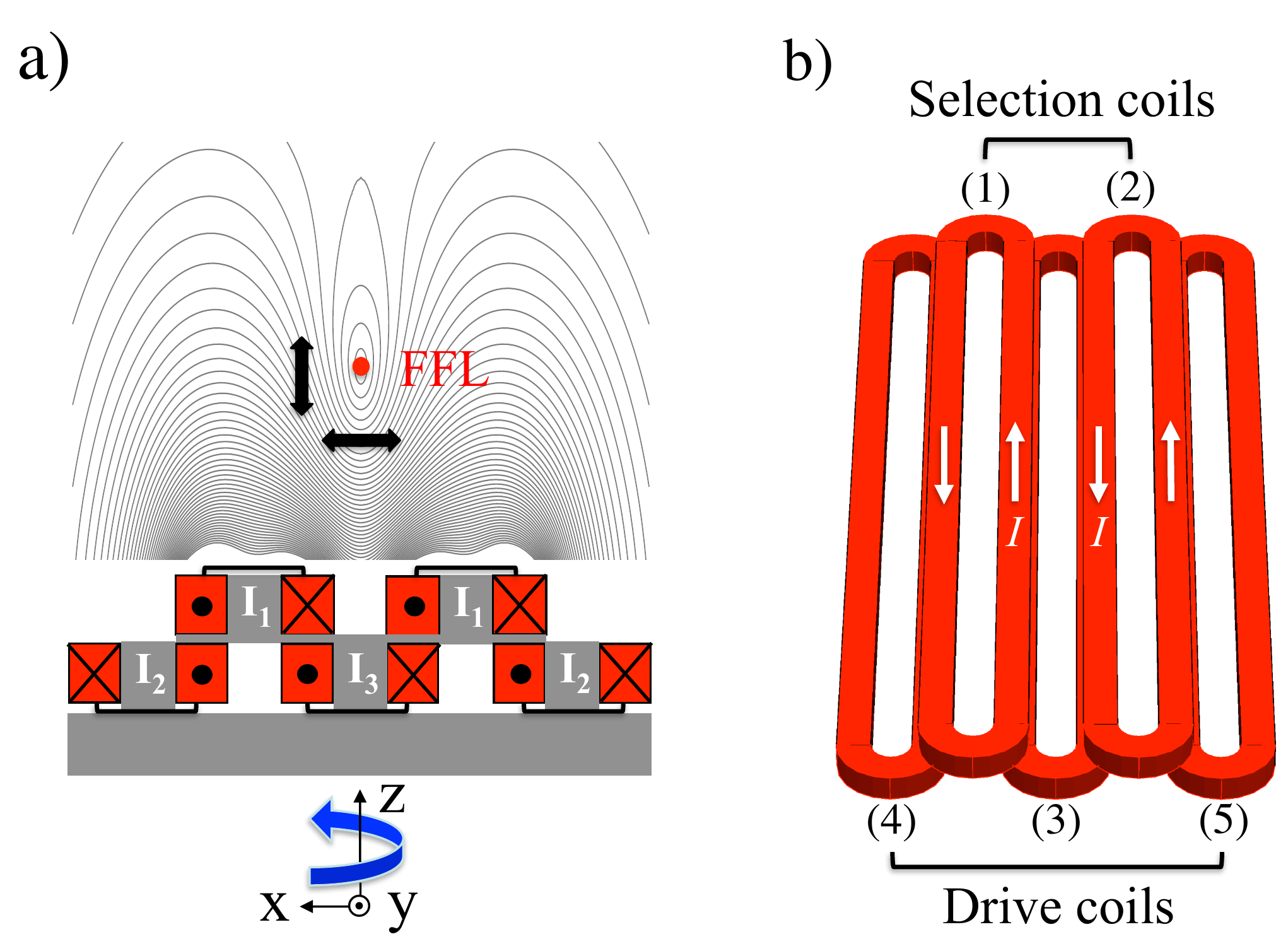}
 \caption{(a) A diagram of single-sided FFL Selection-Drive coils,  
the {\em xz}-contour plot of the magnetic field shows the FFL above the coil's surface generated by a pair of top coils; 
(b) a simulation model of the five-element elongated coil structure. 
To oscillate the FFL along the {\em z}- or {\em x}- axes, as shown by the black arrows, the AC current is applied in the coil (3) or the pair of the coils (4,5), respectively. The setup is mounted on a gantry that is mechanically rotated around the {\em z}-axis enabling in-plane projection encoding. }
 \label{setup}
 \end{figure}
\begin{figure}[tb!]
 \centering
 \includegraphics[width=2.4 in]{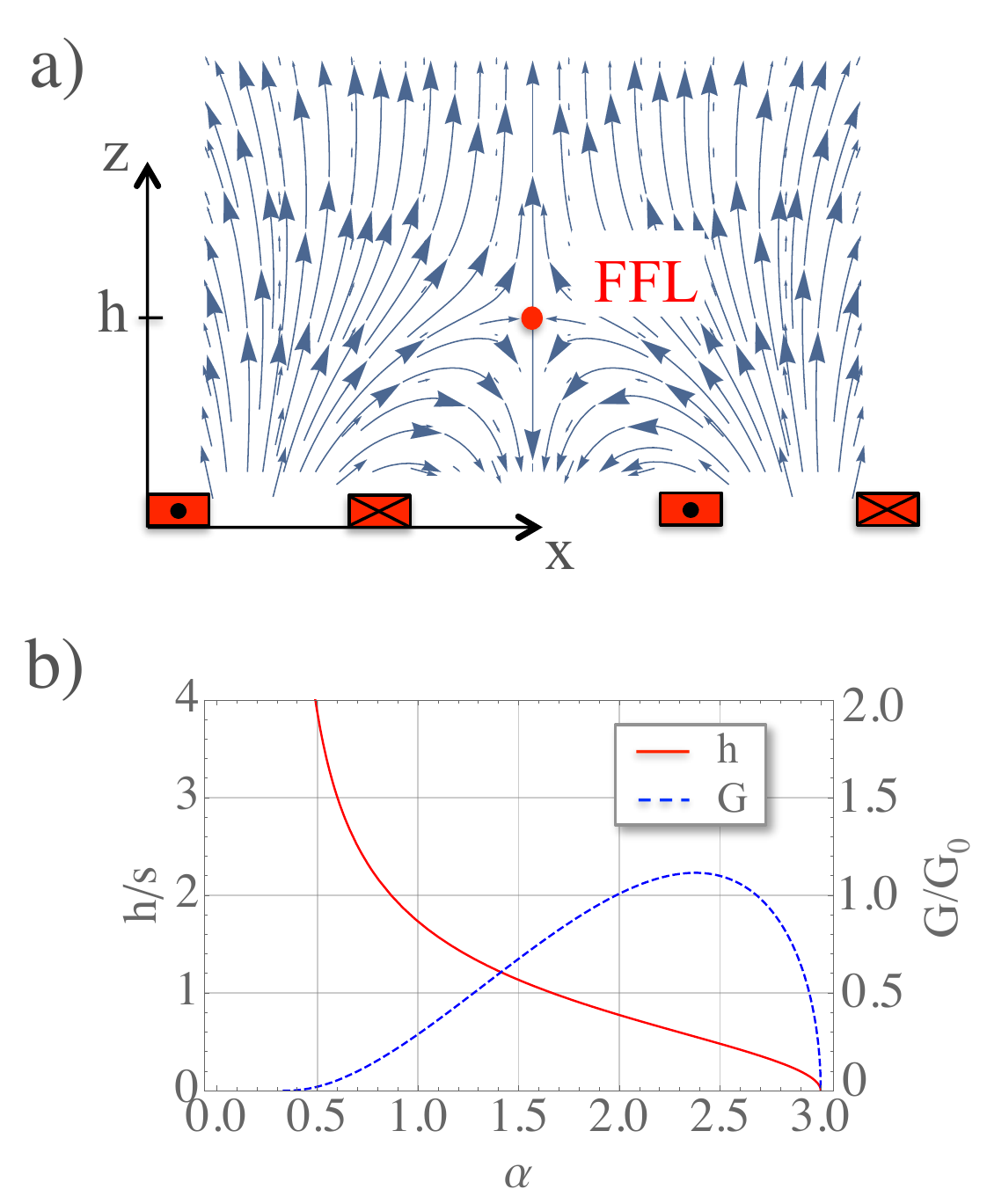}
 \caption{(a) Vector plot of the static magnetic field in the {\em xz-}plane generated by four conductors; (b) a normalized height (the red solid line) and a gradient (the blue dashed line) of an FFL above the surface as a function of the current ratio $\alpha$ for equally spaced conductors.}
 \label{vector}
 \end{figure}

%
The apparatus and the topology of FFL-based selection coils are shown in Fig.~\ref{setup}. In general such selection coils consist of several elongated coils placed in parallel to each other. 
To create a static FFL at some height $z^{FFL}= h$ from the surface of the coils only the two coils are required (top coils (1,2) in Fig.~\ref{setup}). To understand the principle of FFL generation above the surface let us consider four parallel infinite conductors. The superposition of the magnetic fields from these conductors is shown in Fig.~\ref{vector}(a). 
The two innermost conductors with the equal counter-propagating current create ${\bf B}_{in}$ field, which is oriented along 
$\bf -\hat{z}$ in the isoplane ($x=0$). In order to get a field zero line in that plane an additional bias field ${\bf B}_{out}$ is required. Such a field oriented along 
$\bf \hat{z}$ can be created by two outer conductors with the equal DC counter-propagating current~\cite{Dekker00}. 
If we assume infinite length conductors and neglect their finite cross-sections, then from Biot-Savart law the total $B_z$ field is given by

\begin{equation}
B_z(Z,x=0)=B_0\bigg\{ \frac{\alpha\beta}{\beta^2+Z^2} -\frac{1}{1+Z^2} \bigg\} , 
\label{field}
\end{equation}
where $B_0= \mu_0 I_1^{in}/2\pi s$, $\mu_0$ is the permeability of free space, $\alpha = I_1^{out}/I_1^{in}$ is the ratio of the currents in the pairs of the outer and the inner conductors, $\beta = 2d/2s$ is the ratio of the separations of the outer $2d$ and inner $2s$ conductors.
By setting $B_z=0$ in Eq.~\ref{field} we can obtain the height of the FFL as a function of $\alpha$ to be
\begin{equation}
h(\alpha)=\pm s \sqrt{\frac{\beta-\alpha}{\alpha-1/\beta}}.
\label{height}
\end{equation}
%
Eq.~\ref{height} defines the critical current ratio $\alpha =\beta$ so that $\beta^{-1} I_1^{in} < I_1^{out} < \beta I_1^{in}$ for the FFL to be above the surface at a finite height. 
The gradient of the magnetic field at $z=h$ is given by

\begin{equation}
G_{zz}(\alpha)=G_0 \frac{4}{\alpha (\beta^2-1)^2} \sqrt{ (1-\alpha/\beta )(\alpha \beta-1)^5} , 
\label{gradient}
\end{equation}
where $G_0=B_0/s$.
The gradient can be maximized for a certain current ratio given by 
\begin{equation}
\alpha_{max}=\lambda + \sqrt{ \lambda+1/2} , 
\label{maximum}
\end{equation}
where $\lambda=(3\beta^2-1)/8\beta$. When the conductors are equally spaced so that $\beta=3$ then from Eq.~\ref{gradient} the maximum gradient for the given $I_1^{in}$ is 
$G_{zz}^{max} \approx 1.1 G_0 $ for $\alpha_{max}=2.38$. Figure~\ref{vector}(b) shows the dependence of the normalized height of FFL and the normalized gradient vs current ratio 
$\alpha$.

Different from the ideal four-wire geometry for the elongated coils, the current $I_1$ must be the same in all four conductors, {\em i.e.} $I_1^{out} =I_1^{in}$, thus the FFL height {\em h} is independent on the current amplitude. As shown in Fig.~\ref{vector}(b) the maximum gradient cannot be reached at $\alpha=1$ and an independent current control is required that can be achieved by the additional coil. 
From Eq.~\ref{height} setting $\alpha =1$ we obtain the position of FFL above the surface to be $h = 1.7s$, which does not take into account the finite cross-sections of the conductors.  
We note that, the finite length of the coils contributes to the curvature of the FFL along $\bf \hat{y}$.

\subsection{Dynamic FFL operation for spatial encoding}
\label{operation}
While two elongated coils are necessary and sufficient to generate a stationary FFL at a fixed height, for MPI it is required to repetitively move the FFL in space. 
An arbitrary translation of the FFL in two orthogonal directions {\em x, z} can be accomplished by means of three more elongated coils positioned under the selection coils (1,2). Such a 
three-coil array consists of all equal and equidistant coils with their cores shifted with respect to the top layer by half a separation distance as shown in Fig.~\ref{setup}. We note that the same method of equivalent conductors, in principle, can be applied to the array of five coils positioned in the same top layer so that the coils (4,5) surround coils (1,2) and the latter surround coil (3), however, this configuration is less power efficient due to wider separation per unit conductor size.

Let us first consider the depth encoding by SD coils. A drive field 
${\bf B}^D(t)=B_z^D (t) \bf \hat{z}$ is required in order to shift the height of the FFL in the isoplane
so the total field in the isoplane is: 
\begin{equation}
B_z(z,x=0,t)=G_{zz} z+ B_z^D(t) .
\label{Bz}
\end{equation}
That can be achieved by the drive coil (3), which is located at $x=0$. Since the two long conductors of that coil directly overlap with the inner conductors of the top coils (see Fig.~\ref{setup}(a)), the AC current 

 \begin{figure}[tb!]
 \centering
 \includegraphics[width=2 in]{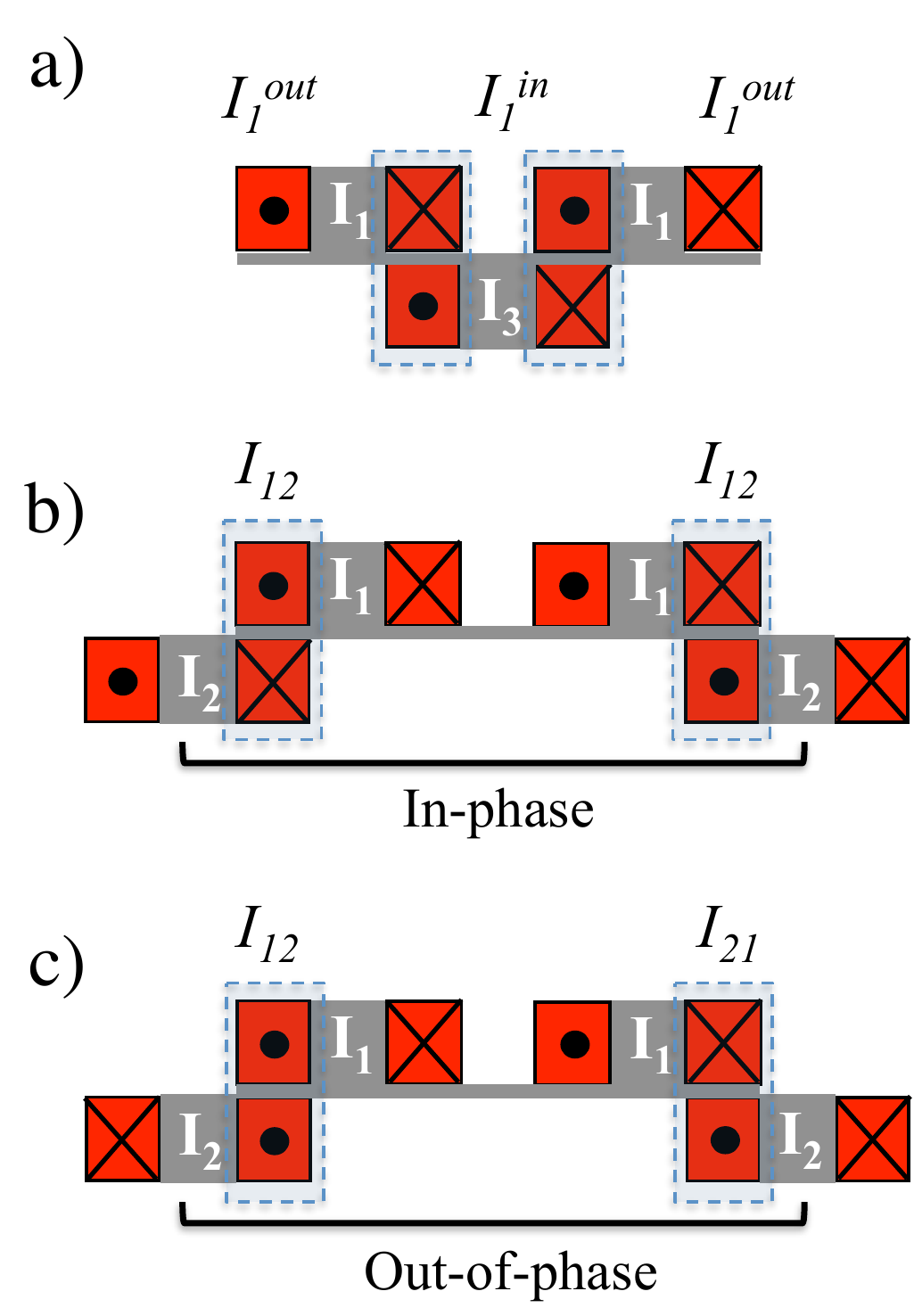}
 \caption{A diagram of four (a) and six (b,c) equivalent conductors composed of an array of the elongated coils. The symmetric arrangement (a,b) allows modulating the height of the FFL in the isoplane, while the arrangement (c) allows shifting the isoplane.}
 \label{current}
 \end{figure}
\begin{equation}
I_3=a_{31} I_1sin(2\pi f_z t ) ,
\label{zdrive}
\end{equation}
where $f_z$ is the drive frequency and $a_{31}$ is the current amplitude ratio, effectively modulates the DC current $I_1$ in the two inner conductors of the selection coils thus breaking the current degeneracy of the elongated loop conductors. Therefore the three-coil array creates an equivalent of the ideal four conductor configuration as shown in Fig.~\ref{current}(a),
which is required to move the FFL in the isoplane with relatively large displacements described by Eq.~\ref{height}. Setting $I^{out}=I_1$ and $I^{in}=I_1\pm |I_3|$ we can apply the theory developed in Sec.~\ref{station}. Thus we get a time-dependent current ratio $\alpha =I_1/(I_1\pm |I_3|)$, from which the critical condition is $I_3 <2/3 I_1$ or $a_{31} < 2/3$. The maximum gradient is obtained for $I_3=0.58 I_1$.
This method, in principle, can provide a fast ($f_z=25$~kHz) and sensitive 1-D MPI along the {\em z}-axis.  

Another way to achieve depth modulation is to utilize the bias field generated by a pair of outer coils (4,5) with AC current $I_2$ in each coil

\begin{equation}
I_2=a_{21} I_1sin(2\pi f t + \phi_i) ,
\label{zdrive2}
\end{equation}
where $f$ is the drive frequency, $a_{21}$ is the current amplitude ratio, and $ \phi_i$ is the phase.
If the current in both coils fed in-phase (see the diagram in Fig.~\ref{current}(b)) then the drive field $\tilde B_z^D(t)$ competes with the selection field generated by the static coils (1,2) according to Eq.~\ref{Bz}. 
Thus Eq.~\ref{field} can be applied for the coils (4,5) with the modified separation ratio $\tilde{\beta}=1+2/\beta=5/3$ and the current ratio $\alpha =1$. It can be shown that only one field zero exists above the surface in this configuration.

The assembly of four-element coil array (1,2,4,5) effectively provides an ideal six-conductor configuration as shown in Fig.~\ref{current}(b). We can estimate the displacement by noticing that for
$a_{21}=1$ we get $I_{12}=0$ (see Fig.~\ref{current}(b)), so the equivalent geometry with four conductors has $\tilde{\beta}=2+\beta=5$. Therefore the positive phase of $I_2$ provides $h=2.7 s$ so the total displacement around the base height is $\pm s$. 

The two methods of the depth encoding described above can be used independently to adjust the gradient of the magnetic field at the different heights and to correct the FFL height during in-plane displacements that are valuable for multidimensional MPI. 

To utilize in-plane projection reconstruction algorithms~\cite{Goodwill12,Konkle13} it is required to oscillate FFL along $\bf \hat{x}$ for each rotation step, which can be done with an external bias field oriented along $\bf \hat{x}$ that effectively shifts the isoplane. This x-drive field can be generated by the pair of the outer coils (4,5) that are driven {\em out-of-phase} as shown in in Fig.~\ref{current}(c). Such an equivalent six-conductor geometry produces asymmetric current pattern $I_{12} \ne I_{21}$ with respect to the isoplane that moves the FFL position along the magnetic field lines $B_x^D(t)$ in the {\em xz}-plane.
The total field is: 
\begin{equation}
B_x(z=h,x,t)=G_{xx} x+ B_x^D(t),
\label{Bx}
\end{equation}
where $G_{xx}=-G_{zz}(h)$. This x-drive field can produce an oscillating FFL according to Eq.~\ref{zdrive2}.   
Thus the in-plane ({\em xy}) encoding can be implemented through a projection imaging by a mechanical rotation of the device or a subject around $\bf \hat{z}$ up to 180$^{\circ}$ while oscillating the FFL along $\bf \hat{x}$. 
Eq.~\ref{Bx} assumes a uniform x-drive field ${\bf B}^D(t) = B_x^D(t) {\bf\hat x}$. This condition, however, holds only for the displacements $x \ll s$. At the larger displacements the x-drive magnetic field lines bend so that $B_z^D(x,t) \ne 0$. That results in the x-dependent FFL height variation $\Delta h=h(x)$. 
For the small field curvatures $\Delta h \approx x^2/(2h_0)$, where $h_0 = h(x=0)$. This height variations can be corrected by applying an additional bias field $\tilde B_z(t)=-B_z^D(t)$, which can be achieved by modulating the current in the coil (3) or the coils (4,5). Thus for known displacements along $\bf \hat{x}$ we can apply Eq.~\ref{height} solving for $\tilde \alpha$ to obtain the modulation current ($I_3$ or $I_2$) amplitude.

The {\em xy}-plane encoding can be combined with the fast or relatively slow depth modulation that can be used for the slice selection.

\section{Materials and methods}
\label{methods}
To evaluate the magnetic field generated by the single-sided elongated coil topology we carried out quasi-static simulations using Wolfram {\em Mathematica\textsuperscript{\textregistered}} software with {\em Radia} package (ESRF, France)~\cite{Radia}. The package allows to calculate flux density $\bf B$ (also called $\bf B$ field here) of a coil by utilizing boundary Integral Methods.

First we consider a static FFL generated by two selection coils (see Fig.~\ref{setup}) as described in Sec.~\ref{station}. For practical considerations, we modeled two identical elongated coils for the small-scale MPI scanner using the properties of the actual coil structure described in~\cite{Tonyushkin10} with the following dimensions: 
$l=200$~mm (straight length), $w= 24$~mm (width) and aspect ratio of $8 : 1$. 
Here, each coil consists of $N=25$ layers of conductor with cross-section 6.33 $\times$ 0.3~mm$^2$, the core is made of a nonconducting material with the width $s=9$~mm, and the coil separation of $\Delta=4s$, which corresponds to $\beta =3$. 

Let us now consider a design of a coil structure that is capable of multidimensional image encoding. The model coil assembly consists of the pair of the selection coils in the upper layer and three identical elongated coils in the lower layer that are shifted along {\em x}-axis by $\Delta /2$ with respect to the selection coils as described in Sec.~\ref{operation} and shown in Fig.~\ref{setup}(b). 

We perform quasi-static field simulations with the reference DC current $I_1=100$~A in the selection coils and the instantaneous current patterns of $I_2, I_3$ corresponding to their positive and negative phases as referenced by the direction of $I_1$. The power consumption of the actual electromagnetic coil at the operating current of 100 A is 2.5 kW per coil. Thus the estimated DC power consumption of the MPI device at the specified gradient strength is 5 kW.

\section{Results and discussion}
\label{results}

\subsection{Static FFL generator}
\label{inplane}
 \begin{figure}[htb!]
  \centering
 \includegraphics[width=3.2 in]{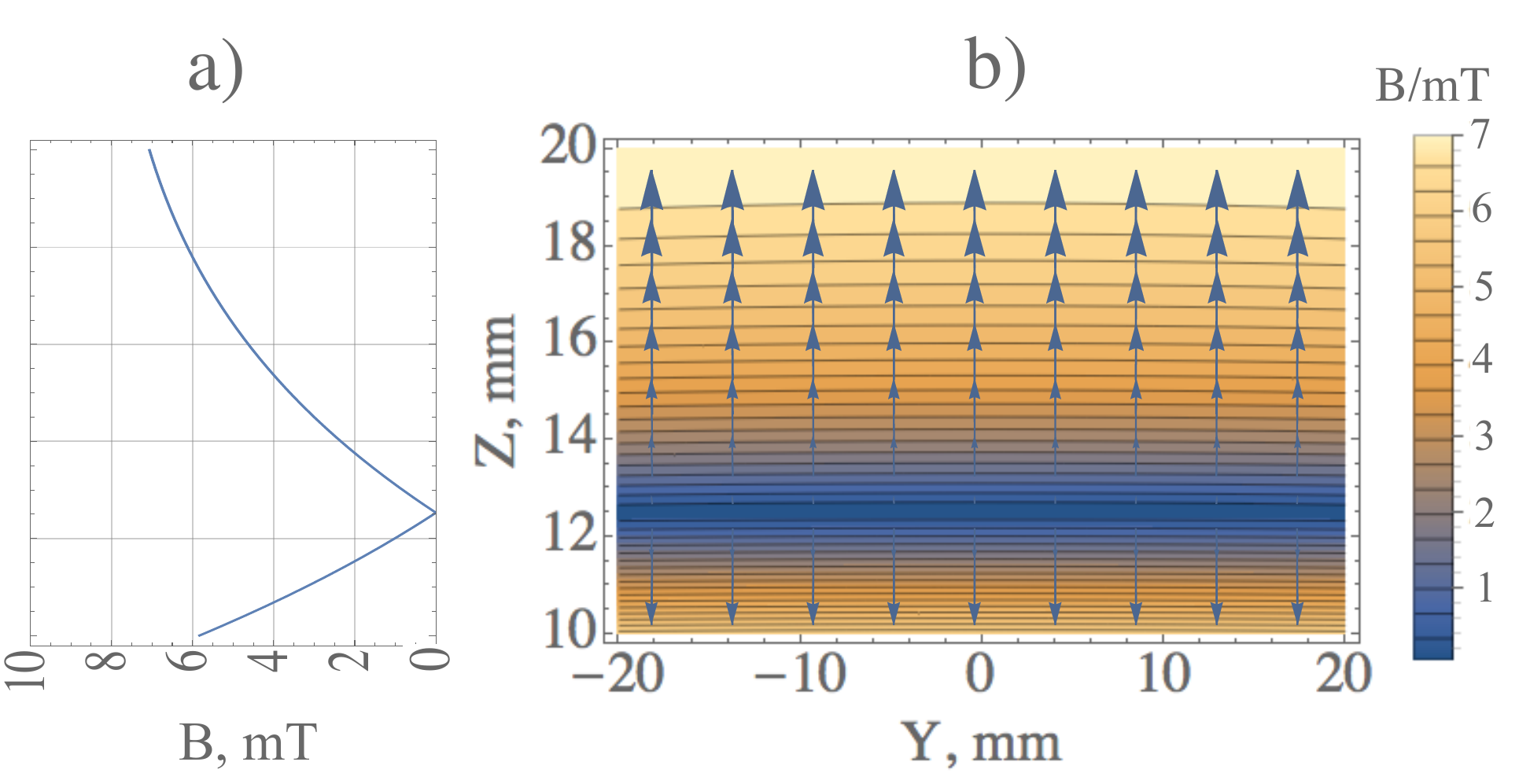}
 \caption{Generation of a static FFL at $z=12.5$~mm above the surface by a pair of selection coils: (a) simulated magnetic field magnitude $B$ along the {\em z}-axis and (b) magnetic field contour plot at the isoplane.}
 \label{fixed}
 \end{figure}
Figure~\ref{fixed} shows the simulation results of the magnetic field $B$ produced by the pair of the selection coils. The simulations were performed for the reference current $I_1$ that generates FFL at $h=12.5$~mm with the field gradient $G_{zz}=2 $~$T/m$, which is $30 \%$ of the maximum gradient as defined by Eq.~\ref{gradient}. 

The practical region of operation along the FFL direction is at least $L \approx 4$~cm without sacrificing resolution as defined by the finite curvature of the magnetic field lines. The curvature of the FFL depends on the aspect ratio of the coils and will be further discussed in Sec.~\ref{discuss}. 

\subsection{FFL generator for depth encoding}
\label{3D}
%
 \begin{figure}[tb!]
  \centering
 \includegraphics[width=3.2 in]{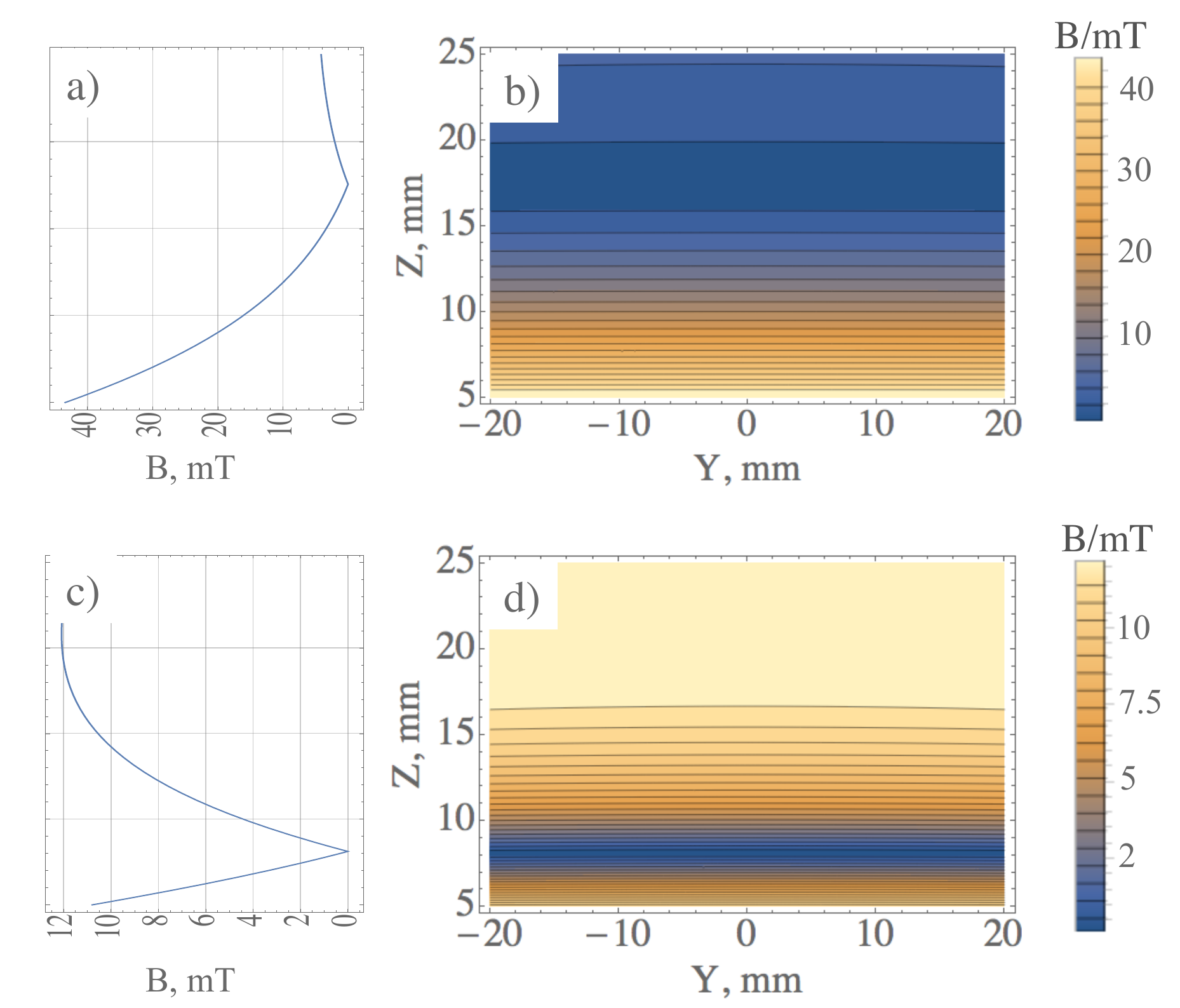}
 \caption{Depth encoding with the single-sided MPI device: (a,c) simulated magnetic field magnitude $B$ at the isoplane along the {\em z}-axis and (b,d) magnetic field {\em yz}-contour plots show the FFL at (a,b) $z=17.5$~mm ($I_3=-0.5I_1$) and (c,d) $z=8$~mm ($I_3=0.5I_1$) above the surface, respectively.}
 \label{depth}
 \end{figure}

Figure~\ref{depth} shows simulation results of the depth modulation by means of the drive coil (3) (see Fig.~\ref{setup}(b)). In the model we fixed the current $I_1$ in the selection coils and vary $I_3= \pm 0.5I_1$ in the drive coil, so that $\alpha =$2 and 0.63 for the positive and negative phases of $I_3$, respectively. Figs.~\ref{depth}(a,b) show the magnetic field for $I_3=-0.5I_1$ with the FFL height at 17.5 mm above the surface and the gradient $G_{zz}=1.3$~$T/m$, and Figs.~\ref{depth}(c,d) show the field patterns for $I_2=0.5I_1$ with the FFL height at 8 mm above the surface and the gradient $G_{zz}=2$~$T/m$. Thus the depth FOV spans $\Delta h \approx 1$~cm. Due to the inhomogeneity of the field in the asymmetric topology the gradient strength varies with the height that affects the resolution of MPI, therefore current adjustments are necessary. Since the field gradient $G_{zz}$ is proportional to $I_1$ (see Eq.~\ref{gradient}) it is possible to correct the gradient by dynamically adjusting $I_1$ while keeping $\alpha=const$. 
This modulation algorithm can be implemented by substituting $I_1$ in Eq.~\ref{zdrive} with the current $\tilde I_1$ modulated by the triangular wave according to:

\begin{equation}
\tilde I_1(t)= \frac{2}{5}I_1\left(4-\frac{3}{T} \left| t-\frac{T}{4} \right| \right)  , 
\label{modulation}
\end{equation}
for $-T/4<t<3T/4$ and periodic with the period $T=1/f$. 
Thus at $h=17.5$~mm $\tilde I_1=1.6I_1$ and at $h=8$~mm $\tilde I_1=I_1$ so the gradient $G_{zz}=2$~T/m at the boundaries of the FOV. 

\subsection{FFL generator for in-plane encoding}
\label{3D}
%
 \begin{figure}[tb!]
 \centering
 \includegraphics[width=2.4 in]{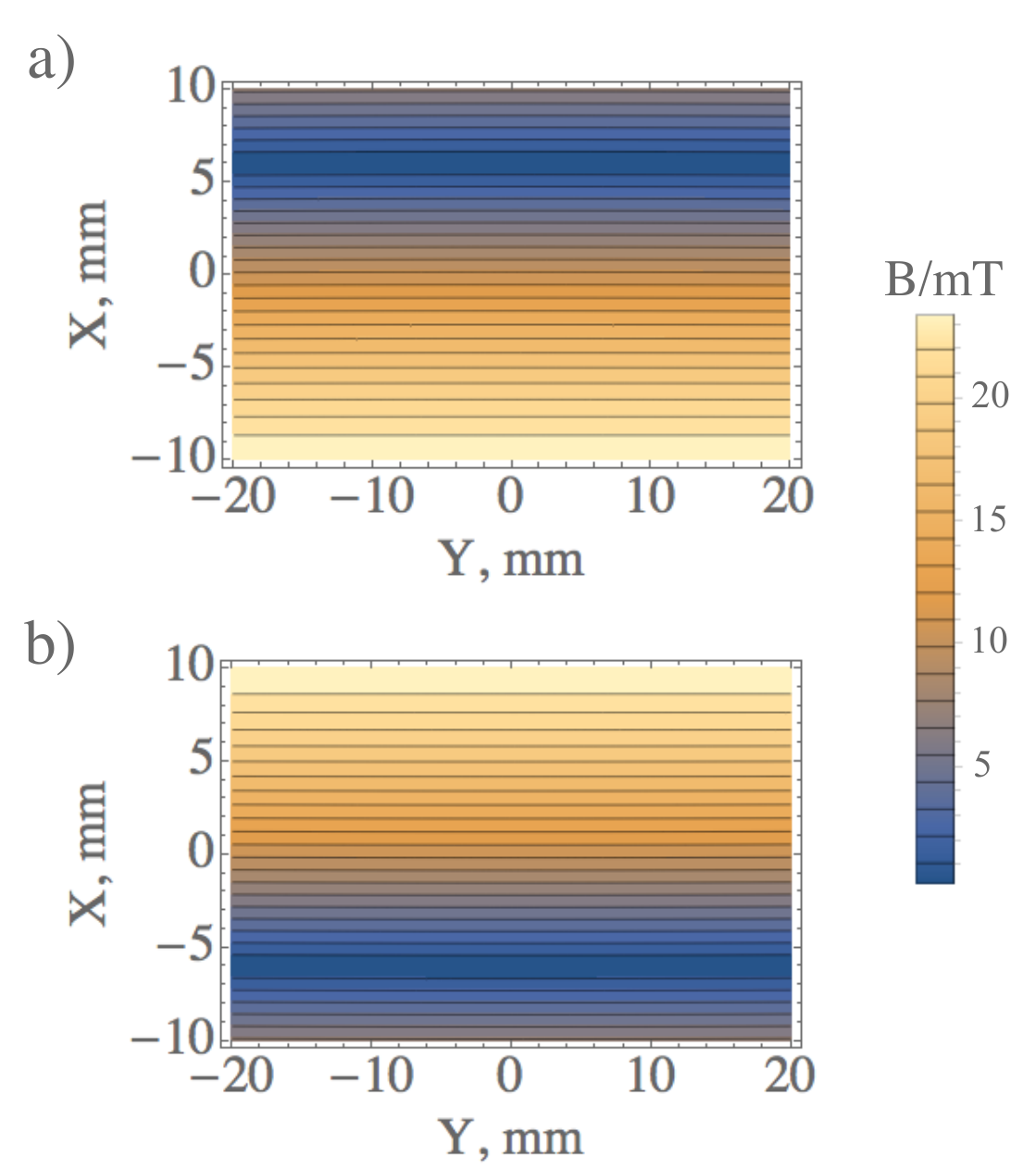}
 \caption{In-plane encoding: (a,b) simulated magnetic field contour plots in the {\em xy}-plane and $h=14.2$~mm show the FFL spans $\Delta x =\pm 6$~mm for the current: 
  $I_3=-0.5I_1$ and (a) $I_2=I_1$, (b) $I_2=-I_1$. }
 \label{xyplane}
 \end{figure}
%
 \begin{figure*}[htb!]
 \centering
 \includegraphics[width=5.4 in]{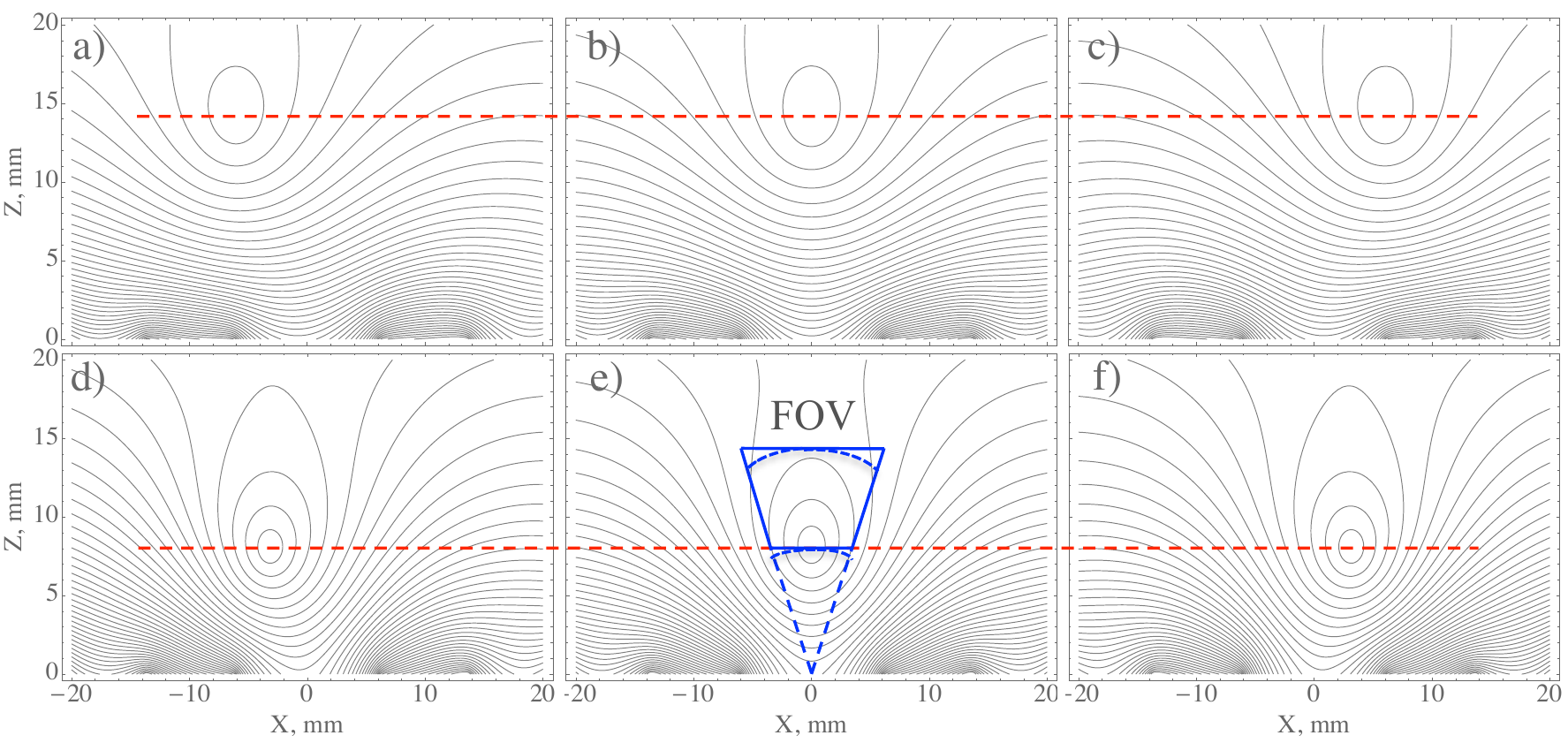}
 \caption{Simulated magnetic field contour plots in the {\em xz}-plane show the FFL spans: (a-c) $\Delta x =\pm 6$~mm at $h=14.2$~mm, (d-f) $\Delta x = \pm 3$~mm at $h=8$~mm. The horizontal red dashed lines show in-plane constant height trajectory, (e) the blue dashed and solid lines define the FOV before and after the height corrections due to the curved field lines, respectively. The current:  (a) $I_3=-0.5I_1$, $I_2=-I_1$, (b) $I_3=-0.2I_1$, $I_2=0$, (c) $I_3=-0.5I_1$, $I_2=I_1$, (d) $I_3=0.4I_1$, $I_2=-I_1$, (e) $I_3=0.5I_1$, $I_2=0$, and (f) $I_3=0.4$, $I_2=I_1$. }
 \label{xzplane}
 \end{figure*}
The set of the drive coils (4,5) can be used for the FFL translation along $\bf \hat{x}$ for in-plane ($z=h$) spatial encoding. In this operation regime the x-drive coils are fed with an out of phase AC current $I_2$ according to the pattern shown in Fig.~\ref{current}(c). Figure~\ref{xyplane} shows simulation results of the magnetic field pattern. 
The presented examples show in-plane ($z=14.2$~mm) FFL total displacement of $\Delta x= 1.2$~cm for the x-drive current  $I_2= \pm 0.5I_1$ and the additional z-drive current of $I_3=-0.5I_1$. 

As was pointed out before, the FFL trajectory in the {\em xz}-plane is not linear therefore for the cartesian encoding scheme an independent FFL height adjustment is required. Such an adjustment can be achieved either by the z-drive coil (3) or by the same coils (4,5) with the in-phase modulation of $I_2$. Figure~\ref{xzplane} shows the simulated magnetic field contour plots in the {\em xz}-plane that represent the adjusted constant-height trajectory from left to right due to the modulation current $I_3$ in the coil (3). 
 
 \subsection{Discussion}
 \label{discuss}
The simulation results demonstrate, that the array of the electromagnetic coils in combination with the rotation around $\bf \hat{z}$ can spatially encode the cylindrical FOV with the diameter that is limited by the usable FFL length and the height $\Delta h \approx 1$~cm, while keeping the field gradient constant. The {\em xy}-plane FOV can be significantly increased with the introduction of the x-focus coils. The operation of the focus coils is following: the first partial FOV (pFOV) is created by the original arrangement with the selection coils (1,2), the second pFOV is shifted to the left (see Fig.~\ref{setup}(b)) by using the coils (3,4) as the selection coils and the coil (1) as the z-drive coil, and the third pFOV is similarly shifted to the right by using the coils (3,5) as the selection coils and the coil (2) as the z-drive coil. Adding two more coils in the top layer to surround coils (1,2) would allow 
the x-drive operation within each pFOV.

 \begin{figure}[tb!]
  \centering
 \includegraphics[width=2.8 in]{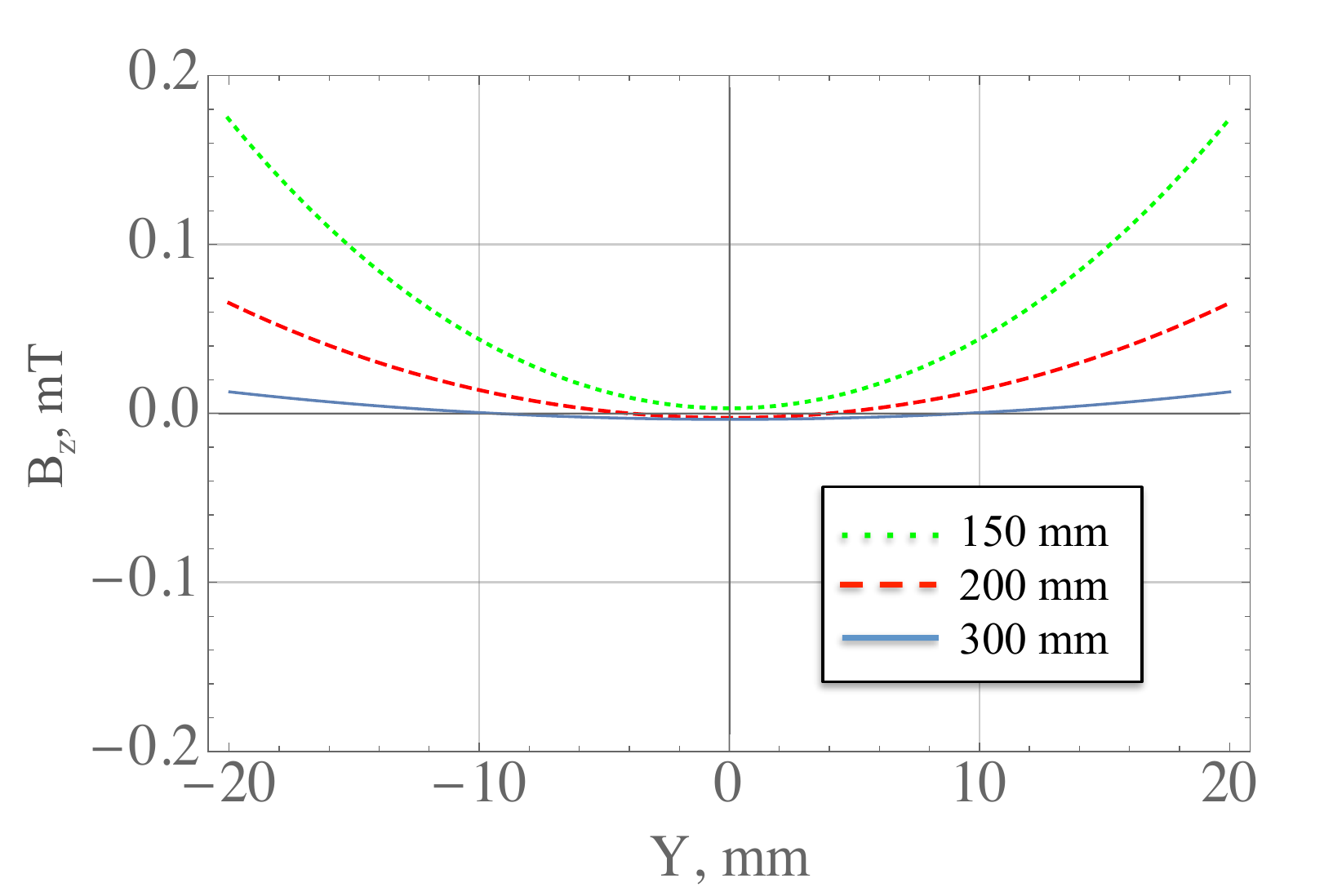}
 \caption{$B_z$ field as a function of the coordinate {\em y} at $h=12.5$~mm and $x=0$ shows the curvature of the FFL versus various lengths $l$ of the coils.}
 \label{curvature}
 \end{figure}
One of the features of the proposed design is a low longitudinal curvature of the field along $\bf \hat{y}$. The curved magnetic field lines (see contour plots in Figs.~\ref{fixed},\ref{depth}) arise due to the finite size of each coil. 
Figure~\ref{curvature} shows the simulation study of the FFL curvature versus the length of the coils. 
As seen from the simulations, for the coils with length $l \ge 200$~mm, the divergence of the equipotential lines is less than $100 \, \mu$T (at $I_1=100$~A) per 2 cm length of the FFL, which is tolerable for the practical MPI device~\cite{Goodwill12}. 
A usable FFL region can be selected by utilizing the spatial sensitivity profiles of the receiving coils. Such elongated receive coils can be placed on top of the selection coils assembly and in parallel with the respective drive coils.

The geometry with several pairs of electromagnets can potentially generate more than a single FFL at the same time. However,  those secondary zeroes are located either near the surface or far outside of the encoded FOV and thus can be excluded either by proper positioning the object or by utilizing the narrow spatial sensitivity profile of the receiving coils.

\section{Conclusion and Outlook}

We presented a novel design, a method, and field correction algorithms for a single-sided FFL-based device operation, which is capable of multidimensional imaging. Such a device, once built, can provide a relatively large FOV with no constraints on the object's size and a magnetic field gradient with a flat FFL.  The presented model coils can encode the whole volume of a small animal or penetrate deep enough into human organs such as in the vascular or lymphatic systems. An MPI device based on the proposed selection field assembly can be more compact and robust in comparison to the state-of-the-art FFP-based MPI scanners.

A modified low power consumption MPI device can be constructed with a hybrid system, where the pair of the inner selection coils can in principle be replaced by two pairs of elongated vertically polled permanent magnets \cite{Tonyushkin} or ferromagnets thus relaxing cooling requirements and reducing power consumption by at least a factor of 100. 
The excitation and slice encoding can be done with the outer electromagnets following the method proposed here.

\ack
We acknowledge support from University of Massachusetts President Office through OTCV Award.

\end{document}